\documentclass[twocolumn]{revtex4-1} 
\usepackage{graphicx}
\usepackage{subfig}
\usepackage{color}
\usepackage{epstopdf}
\usepackage{amsfonts}
\usepackage{amsmath}

\begin{document}

\title{Morphological properties of 2D symmetric Airy beams extracted from the stationary wave approximation}
\author{F. Camas-Aquino}
\affiliation{Instituto de F\'{\i}sica, Universidad Nacional Aut\'onoma de M\'exico}
\address{Apartado Postal 20-364, 01000, Cd. de M\'exico, M\'exico.}
\author{P. A. Quinto-Su}
\affiliation{Instituto de Ciencias Nucleares, Universidad Nacional Aut\'onoma de M\'exico}
\address{Apartado Postal 70-543, 04510, Cd. de M\'exico, M\'exico.}
\author{ R. J\'auregui}
\email{rocio@fisica.unam.mx}
\affiliation{Instituto de F\'{\i}sica, Universidad Nacional Aut\'onoma de M\'exico}
\address{Apartado Postal 20-364, 01000, Cd. de M\'exico, M\'exico.}

\begin{abstract}
We explore the morphological properties of symmetric Airy beams in the paraxial and nonparaxial regimes. We consider a 2D electromagnetic realization with a single transverse component of the electric field, and in the nonparaxial regime, the longitudinal component along the optic axis. The general structure of these beams is analyzed with the combination of several approaches: geometrical optics through the use of caustics,  the asymptotic wave properties of the light field using the stationary wave approximation and numerical integration. The geometrical optics approach involves locating the critical points that are later used in the stationary phase approximation. In the paraxial regime the highest order of the roots is 3, while in the nonparaxial regime, the order can be of up to 6. The technique yields conditions to identify interesting features on the beam, like the number of waves interfering constructively/destructively at the critical positions. The results are confirmed by the numerical simulations. In this way it is possible to distinguish and classify phase singularities like optical vortices and dislocations. The developed algorithm could be used to study any structured light field.

\end{abstract}

\maketitle

\section{Introduction}

The analysis of structured light fields using concepts developed in Singular Optics \cite{Soskin} allows a greater understanding and yields to the identification of potential applications of such fields. For instance, strong variations of optical fields around phase singularities enhance the sensitivity to small changes in the propagation medium. This characteristic is useful for the manipulation of internal and external motion of microparticles \cite{dholakia} and atoms \cite{phillips}, it is also relevant for super-resolution imaging \cite{imaging1, imaging2, imaging3} and spectroscopy \cite{RJ1,RJ2,Schmiegelow,spec}, and it has important implications to solar energy harvesting \cite{harv1,harv2}, and astronomical observations \cite{teles}.

In this work, we perform a study of the singularities of   Airy symmetric (AS) electromagnetic (EM) waves in and out the paraxial regime, using both an undulatory approach and a geometric optics analysis. Airy symmetric beams are also known as Airy-Scorer beams \cite{Vaveliuk1,Quinto,Jauregui,Vaveliuk}. The latter name  arises from the relation of their mathematical expression to the Airy and the Scorer functions. AS beams result from imposing to Airy beams \cite{berryaj,Siviloglou1,Siviloglou2} a symmetry under reflection of the transverse coordinate.

Experimentally, AS beams can be generated by imprinting the proper phase to an expanded Gaussian beam using a spatial light modulator, as it has been reported in Refs.~\cite{Vaveliuk1,Jauregui,Quinto}. Focusing of a 3 dimensional AS electromagnetic field into well defined spots with a pyramidal geometry, gives rise to large gradients of the intensity that have shown to be useful for optical trapping, and their structural stability has been experimentally tested by blocking parts in its optical path \cite{Quinto}.

In the short-wave limit, a geometric study can be used to identify  relevant  structures of any EM  wave. For example, from the  geometric optics approach arises the concept of caustics: which are theoretical singularities formed by the coalescence of multiple rays that lead to high intensity regions. Besides, caustics separate regions with a discontinuity in the number of rays that cross each point in space. Most Geometric Optics studies are carried out for EM waves that can be described within the paraxial approximation. Outside such a regime, the expression that defines the rays is modified and subtle differences with respect to paraxial systems may arise.

Similarly to Airy beams, the angular spectra of AS beams depend on the third power of the transverse components of the wavevector, but via an absolute value. As a consequence, the morphology of the caustics does not correspond to the generic classification of diffraction catastrophe theory in terms of a polynomial \cite{Berry3,Arnold2,Nye1,Arnold1}. The structure of two dimensional (2D) AS beams is similar to that of the Pearcey beams \cite{Pearcey}, {\it i.e.}, to the cusp diffraction catastrophe \cite{Berry3}.

Undulatory studies usually take into account that any EM wave can  be written as a superposition of elementary plane waves via their frequency and angular spectra. Notice that an interesting link between the undulatory and geometric analyses results when, in the neighbourhood of a spatial location where a striking feature of the EM field is expected, a finite set of plane waves are identified whose superposition reproduces the general characteristics of such a feature. Along this line of thought, optical vortex lines can be formed by the interference of at least three plane waves \cite{Braunbek}. The number of waves is a relevant parameter to define possible vortex lines topologies \cite{DennisHolleran}.

This manuscript reports a careful study of the morphology of AS beams. In Section II, the notation and general concepts that are used for the analysis of the AS beams are introduced including the definition of some related beams. The numerical method to characterize the beams is described in Section III. Section IV is devoted to 2D AS beams.  In this Section, previous studies of the distribution of maxima and minima of the intensity are revisited \cite{Vaveliuk}. The nonparaxial case is studied in section V. Conclusions of this work are presented in the last Section.

\section{General structure of AS beams and other related electromagnetic fields.}

Before working on the AS beams we set some general concepts and examples through their application to Airy and Pearcey beams.

A monochromatic scalar wave $\psi(x_1,x_2,x_3)\mathrm{e}^{-i\omega t}$ of frequency $\omega$ and velocity in vacuum $c$, can be characterized by its angular spectrum $\mathfrak{A}(k_{s_1},k_{s_2})$,
\begin{eqnarray}
\psi_\omega(x_1,x_2,x_3) &=& \int dk_1\int dk_2 \vert\mathfrak{A}(k_{s_1},k_{s_2})\vert\\ 
\nonumber
&\times & \mathrm{e}^{i\Phi(k_{s_1},k_{s_2};s_1,s_2,s_3)},
\label{eq:angspec}
\end{eqnarray}
where, given a length scale $x_0$, $s_{1,2,3} =x_{1,2,3}/x_0$ are  the dimensionless  coordinate  components, $k_{s_{1,2,3}}= k_{1,2,3} x_0$ refer to the dimensionless wavevector components and $k_s =\sqrt{k_{s_1}^2 +k_{s_2}^2+k_{s_3}^2}$ to its magnitude. In Eq.~(1), $\Phi(k_{s_1},k_{s_2};s_1,s_2,s_3) =k_{s_{1}}s_{1}+k_{s_2}s_2 + k_{s_3}s_3 + \phi_{\mathfrak{A}}$ denotes the overall phase and $\phi_{\mathfrak{A}}$ denotes the phase of $\mathfrak{A}$. For monochromatic light waves  in linear isotropic media, the integral in Eq.~(1) must be performed under the condition $\vert {\bf k}\vert= n(\omega)\omega/c =k$, with $n(\omega)$ the refractive index. In this work, we focus on superpositions involving just propagating waves so that $\bf k$ is taken as a real vector. A brief discussion on possible effects associated to the presence of evanescent terms is presented in the last Section.

In the paraxial regime, the involved wavevectors ${\mathbf{k}}$ are such that, for a given direction, here taken to be the third one, the transverse components $k_1$ and $k_2$ satisfy  the condition $\vert k_1 \vert,\vert k_2 \vert \ll \vert k_3\vert \sim  n(\omega)\omega/c$. Then $\psi_\omega(x_1,x_2,x_3)$ is approximated by the function 
\begin{eqnarray}
\psi^{px}_\omega(x_1,x_2,x_3) &=& \mathrm{e}^{ik_s s_3}\int dk_1\int dk_2 \mathfrak{A}(k_{s_1},k_{s_2}) \\ \nonumber
& \times & \mathrm{e}^{i(k_{s_{1}}s_{1}+k_{s_{2}}s_{2}-(k_{s_{1}}^{2}+k_{s_2}^2)\zeta/2)},
\end{eqnarray} 
with $\zeta =x_3c/n(\omega)\omega x_0^2$. The paraxial condition can be mathematically implemented by the proper selection of the integration region. This is physically equivalent to select the optical aperture. An alternative is to  choose an angular spectrum that is not negligible only on the appropriate wavevector region.

The bidimensional finite energy Airy beam \cite{Siviloglou1} is usually defined by the following angular spectrum:
\begin{equation}
\mathfrak{F}_{a_{1}}^{(1)}(k_{s_{1}})=\mathrm{e}^{(a_{1}-ik_{s_{1}})^{3}/3}\delta( k_{s_2}), 
\label{eq:Airy}
\end{equation}
with $\delta(k_{s_2})$ the Dirac delta-function. The positive factor $a_{1}$ gives rise to a Gaussian envelope that can be used to guarantee the validity of the paraxial approximation for a given frequency of the wave in a given medium; the paraxial condition requires $\sqrt{a_1} k_s =\sqrt{a_1}(\omega n(\omega) x_0/c)\gg 1$. The overall phase in ${\bf k}$ space  for a finite energy Airy beam in the paraxial regime is
\begin{equation}
\Phi^{px}_\mathfrak{F}(k_{s_{1}};s_1,\zeta) = k_s s_3+ k_{s_{1}}^{3}/3 - a_{1}^2k_{s_{1}} +k_{s_{1}}s_{1}-k_{s_1}^2\zeta/2,
\end{equation} 
while out of this regime 
\begin{equation}
\Phi^{npx}_\mathfrak{F}(k_{s_{1}};s_1,s_3) =k_{s_{1}}^{3}/3 - a_{1}^2k_{s_{1}} +k_{s_1}s_1 + \sqrt{k_s^2 -k^2_{s_{1}}} s_3.
\end{equation}
Notice the presence of the parameter $a_1$ in the expressions of the overall phase.

Similarly, the bidimensional finite energy Pearcey beam could also be defined in terms of the angular spectrum 
\begin{equation}
\mathfrak{P}_{a_{1}}^{(1)}(k_{s_{1}})=\mathrm{e}^{ik_{s_{1}}^{4}/4-a_{1}k_{s_{1}}^{2}}\delta(k_{s_2}) 
\label{eq:Pearcey}
\end{equation}
yielding an overall phase
\begin{equation}
\Phi^{px}_\mathfrak{P}(k_{s_{1}};s_1,\zeta) = k_s s_3 +
k_{s_{1}}^{4}/4  +k_{s_{1}}s_{1}-k_{s_1}^2\zeta/2
\end{equation} 
in the paraxial regime, and
\begin{equation}
\Phi^{npx}_\mathfrak{P}(k_{s_{1}};s_1,s_3) =
k_{s_{1}}^{4}/4  +k_{s_1}s_1 + \sqrt{k_s^2 -k^2_{s_{1}}} s_3
\end{equation} 
in the nonparaxial regime.

The  Airy, Eq.~(\ref{eq:Airy}), and Pearcey, Eq.~(\ref{eq:Pearcey}), angular spectra give rise to perhaps the simplest waves that exhibit stable caustics. In general, the equations representing rays are obtained by setting to zero first order derivatives of $\Phi$ with respect to the independent components of the wavevector
\begin{equation}
\frac{\partial\Phi}{\partial k_{s_i}} = 0
\label{eq:2Dray};
\end{equation}
caustic curves are obtained by setting to zero both the first and the second order partial derivatives of the overall phase with respect to those components. For the bidimensional beams, the conditions that define caustics are
\begin{eqnarray}
\frac{\partial\Phi}{\partial k_{s_1}} &=& 0,
\label{eq:2Dcaus1}\\
\frac{\partial^2\Phi}{\partial k_{s_1}^2} &=& 0.
\label{eq:2Dcaus2}
\end{eqnarray}
In the paraxial regime, these equations do not depend on the beam frequency.

For Airy and Pearcey paraxial beams the caustic equations are $$s_{1}-a_{1}^{2}-\zeta^{2}/4=0$$ and $$s_{1}^{2}-4\zeta^{3}/27=0,$$ respectively. They give rise to the fold and cusp diffraction patterns \cite{Berry3,Nye1}. While the Airy caustic equation corresponds to a parabola that extends towards the $s_{1}$ positive axis, has its focus at $s_1=a_1^2$, and   is symmetric about this axis, the caustic equation of a Pearcey beam has a real solution for $s_1$ only if $\zeta\ge 0$. As a result, Pearcey beams have perceptible intensities only on that space region. Airy and Pearcey beams exhibit other interesting properties. Airy beams show self-healing \cite{Broky,Vo2010}, while Pearcey beams show self-healing and autofocusing \cite{Ring}.

Airy-Scorer  beams result from modifying the angular spectrum of Airy beams to impose a symmetric behavior of the wavefunction under reflection of the coordinates transverse to the main direction of propagation \cite{Vaveliuk1,Jauregui}. Their angular spectra $\mathfrak{S}$ depend on the third power of the absolute value of the transverse components of the wavevector, that is, for 2D AS beams,
\begin{equation}
\mathfrak{S}^{(1)}_{a_{1}}(k_{s_{1}}) = \mathrm{e}^{(a_{1} -i\vert k_{s_{1}}\vert )^3/3}\delta( k_{s_2}).
\label{eq:2DAS}
\end{equation}

\subsection{EM fields}

The components of the electric ${\bf E}$ and magnetic field ${\bf B}$ of an EM wave  are solutions of the wave equation and, additionally, satisfy Maxwell equations. The paraxial picture associates a scalar wave $\psi$ to a given component of the electric field $E_a$ in the direction defined by the unit vector $\hat{e}_a$ (which is perpendicular to the unit vector along the main direction of propagation of the beam, $\hat e_{3}$ in this work). The transverse nature of the displacement field 
\begin{equation}
\nabla \cdot {\bf D} = 0 
\label{eq:trans}
\end{equation} 
requires that, in general, EM structured beams exhibit a component along the $\hat{e}_3$ vector. For monochromatic  EM waves that propagate in  linear isotropic local media with permittivity $\epsilon(\omega)$, the condition Eq.~(\ref{eq:trans}) in wavevector space reads $\epsilon(\omega) {\bf k}\cdot \tilde{\bf  E} =0$ which implies
\begin{equation}
\tilde E_3 = -\frac{k_a}{k_3} \tilde E_a
\label{eq:vector}.
\end{equation}

 For paraxial beams, $k_a\ll k_3$, $a=1,2$ so that the component  $E_3$ can be neglected, and a quasi-scalar description based on the field $\psi$ is appropriate. Out of the paraxial regime, taking ${\bf E} = \psi \hat{e}_a + \psi_{long} \hat{e}_3$ by demanding that the angular spectrum of the longitudinal electric field $\psi_{long}$ to satisfy Eq.~(\ref{eq:vector}) the fulfillment of Eq.~(\ref{eq:trans}) is guaranteed. Under this scheme, the associated expression for the magnetic field $\tilde{\bf B}$ in wavevector space  is  $\tilde {\bf B} = ({\bf k_s}/k_s)\times\tilde{\bf E}$.

\section{Critical points and the stationary wave approximation}

The computational methods to evaluate the field amplitude are based on accurate numerical integration of the AS beams expression in configuration space in terms of their angular spectra. For nonparaxial beams the dispersion relation $\vert {\bf k}\vert= n(\omega)\omega/c =k$  with ${\bf k}$ a real vector, is directly imposed in the integral.

In the paraxial regime the integrand and the region of integration should reflect the paraxial condition $k_{s_{1}} \ll k_{s}$. The strategy to implement the paraxial condition adopted in this work, involves the parameter $a_{1}$ that yields an effective cutoff and has to be chosen so that $\sqrt{a_{1}}k_s\gg1$. However $k_s$ does not appear in the paraxial integrand. In this manuscript, the numerical integral in the paraxial regime is evaluated within an interval $(-b,b)$ with several increasing $b$ values until convergence is achieved. For a given value of $a_{1}$, the resulting field properties will be compatible with EM waves only for wave frequencies $\omega$ satisfying the condition $\sqrt{a_{1}}\gg c/(n(\omega)\omega x_0)$.

The method to analyze the beams takes advantage of the complementarity between the geometrical and the undulatory pictures. The former is based on the stationary phase approximation which, in general, for a given N-dimensional integral of a highly oscillatory function,
\begin{eqnarray}
&& \int_{\mathbb{R}^{N}}g(k)e^{i\kappa f(k)}d^Nk \sim \sum _{K^i_{s_1}\in \Gamma }e^{i\kappa f(K^i_{s_1})}e^{\pi i/4\mathrm {sign} (\mathrm {Hess} (f))} \nonumber \\ 
&\times & (2\pi /\kappa)^{N/2}\frac{g(K^i_{s_1})}{\vert\det({\mathrm {Hess} }(f))\vert^{1/2}}+\mathcal{O}(\kappa^{-N/2})
\label{eq:asympt}
\end{eqnarray}
where $\Gamma$ denotes the set critical points $K^i_{s_1}$ which are the roots to the ray equation. At each $s_3$ the stationary wave approximation is performed for all the $s_1$.

The contribution of each root to the approximate representation of the 2D AS beam, Eq.~(\ref{eq:asympt}), depends on the local amplitude $$A(K^i_{s_1})=\frac{g(K^i_{s_1})}{\vert\det({\mathrm {Hess} }(f))\vert^{1/2}}.$$

We can extract the term containing the largest amplitude $A(K^M _{s_1})$ in the sum of the previous equation and factorize it:
\begin{eqnarray}
&& \int_{\mathbb{R}^{N}}g(k)e^{i\kappa f(k)}d^Nk \sim   \\ \nonumber  
&& e^{i u(K^M_{s_1})} A(K^M_{s_1})\left( 1 +\sum _{K^i_{s_1}}e^{i \Delta u(K^M_{s_1}, K^i_{s_1}) }  A_\Delta (K^M_{s_1}, K^i_{s_1})\right) \label{eq:rel}
\end{eqnarray}
where $$u(K^i _{s_1})=\kappa f(K^i _{s_1}) + \pi/4\,\mathrm{sign}(\mathrm{Hess}(f)){\big\rvert}_{ K^i_{s_1}},$$ the relative amplitudes are $$A_\Delta (K^M_{s_1}, K^i_{s_1}) =A(K^i_{s_1})/A(K^M_{s_1}).$$
Whenever the relative phases $\Delta u (K^M_{s_1}, K^i_{s_1})$ satisfy 
\begin{equation}
 \Delta u (K^M_{s_1}, K^i_{s_1}) \equiv u(K^i _{s_1}) - u(K^M _{s_1})= m\pi
\label{eq:constr}\end{equation}
 constructive (destructive) interference between rays $K^i_{s_1}$ and $K^M_{s_1}$ with $m$ an even (odd) integer takes place.

The terms with small relative amplitudes $A_\Delta (K^M_{s_1}, K^i_{s_1})$ can be neglected in this analysis.

The curves resulting from demanding the condition  Eq.~(\ref{eq:constr}) for those phase differences were identified, and an integer was assigned to each point. This integer-- that we denote by $I$ --results from adding a $+1$ (comes from $e^{i\Delta u(K^M_{s_1},K^i_{s_1})}$) for each constructive and a $-1$ for each destructive interference.

In the following sections we concentrate in the AS beams, and make some comparisons with  beams that share some of their properties. We explore both paraxial and 
nonparaxial regimes.

\section{ Two Dimensional Airy Scorer beams}
\subsection{Paraxial 2D AS beam}

The paraxial 2D AS beam spatial function is given by
\begin{equation}
\Psi_{a_{1}}^{(1)}(s_{1},\zeta)=\frac{\mathrm{e}^{ik_s s_3}}{2\pi}\int\limits_{-\infty}^{+\infty}dk_{s_{1}}\mathfrak{S}_{a_{1}}^{(1)}(k_{s_{1}})\mathrm{e}^{i\left(k_{s_{1}}s_{1}-k_{s_{1}}^{2}\zeta/2\right)}.
\end{equation}

The absolute value of the transverse component $k_{s_1}$ in the angular spectrum of 2D AS beams, Eq.~(\ref{eq:2DAS}), yields  its even parity behavior about its reflection with respect to the  $k_{s_3}$ axis, {\it i.e.}, $k_{s_1}\rightarrow - k_{s_1}$. The  Pearcey angular spectrum, Eq.~(\ref{eq:Pearcey}), is also an even function of $k_{s_1}$. The real parameter $a_{1}$ determines the relevant values of $k_{s_1}$ for both beams. We will show that for AS beams, $a_{1}$ also modifies the structure of the caustic. Both Pearcey and 2D AS beams present dark regions nearby the local maximum intensity spots.
Given an $a_1$ value, AS beams exhibit higher peak intensities than Pearcey beams. Figure 1 illustrates the intensity pattern (and other features) of a 2D AS beam.

For $a_1=0$, $\Psi_{a_{1}}^{(1)}(s_{1},\zeta)$ is directly related to the Airy $Ai$ and Scorer $Gi$ functions \cite{Jauregui}.  Notice that, as already mentioned, for a given $a_1$ value the paraxial condition  $k_{s_1}\ll n(\omega) \omega x_0/c$  delimits the  physical realization of  $\Psi_{a_{1}}^{(1)}(s_{1},\zeta)$ to EM waves satisfying $\sqrt{a_{1}}\gg c/(n(\omega)\omega x_0)$.

The overall phase of a  paraxial 2D AS beam is
\begin{equation}
\Phi^{px(1)}_\mathfrak{AS}(k_{s_{1}};s_1,\zeta) = k_s s_3+ \vert k_{s_{1}}\vert^{3}/3 - a_{1}^2\vert k_{s_{1}}\vert +k_{s_{1}}s_{1}-k_{s_1}^2\zeta/2.
\end{equation}
The ray condition, Eq.~(\ref{eq:2Dray}), is equivalent to
\begin{equation}
k_{s_1}^2 \mathrm{sign}(k_{s_1}) - a_{1}^2\mathrm{sign}(k_{s_1}) + s_1 - k_{s_1}\zeta = 0, 
\label{eq:AS1Dpxray}
\end{equation}
while the caustic conditions, Eq.~(\ref{eq:2Dcaus1}-\ref{eq:2Dcaus2}), imply that 
\begin{equation}
|s_{1}|-a_{1}^{2}-\zeta^{2}/4=0,
\label{eq:AS1Dparabolas}
\end{equation} 
if $k_{s_{1}}\neq 0$. In Eq.(\ref{eq:AS1Dpxray}) appears the sign function; for $k_{s_{1}}=0$ its derivative as well as that of the absolute value is not properly defined. Notice that the apex of the parabolas described by Eq.~(\ref{eq:AS1Dparabolas}) is situated at the origin in the limiting case $a_1\rightarrow 0$. For $a_{1}$ not negligible, it is observed that the caustics ``break" at $\zeta=0$ in two branches, and that each branch moves away from the origin with a separation given by $2a_{1}^2$.

\begin{figure}[!h]
\includegraphics[width=9cm]{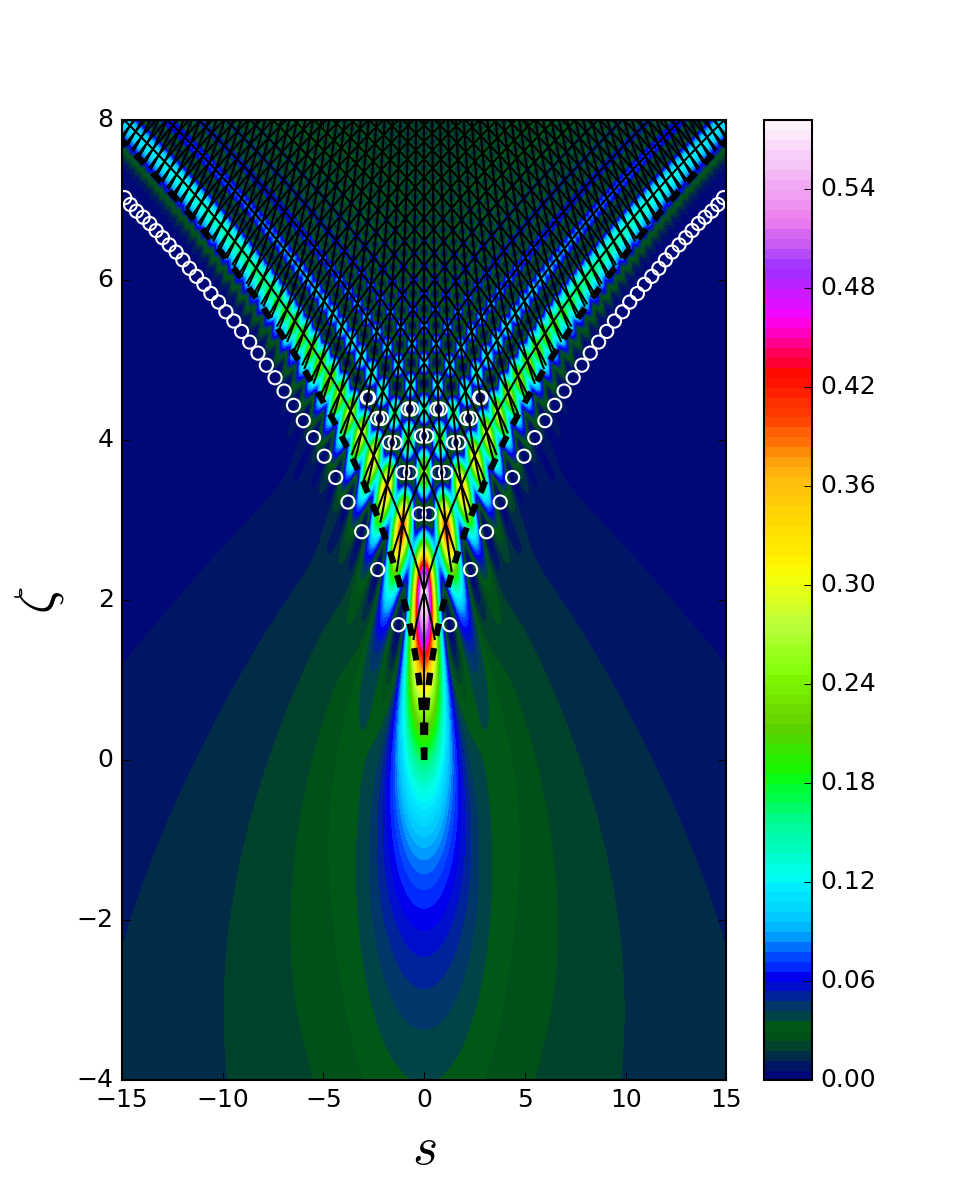}
\caption{Intensity pattern and caustic (black dashed lines) for a paraxial 2D AS beam with $a_{1}=0.03$. Black continuous lines connect the intensity maximums while the minimums are located at the white circles.}
\label{fig:1}
\end{figure}

The caustics of 2D AS beams have a morphology that does not correspond to that expected from the power of the polynomial in the phase of the AS as predicted by standard Catastrophe Optics. That is, 2D AS beams do not exhibit a fold diffraction catastrophe. As expected from the caustic equation, features with parabolic symmetry located on the semiplane with $\zeta>0$ are found. For $a_1\ll 1$, the morphology of the  caustic of a 2D AS beam is a cusp: the symmetry dictates the catastrophe morphology over the expectations from the power dependence of $k_{s_1}$ in the overall phase \cite{Jauregui,Vaveliuk}. In Fig.~\ref{fig:1} the caustics are shown with black dashed lines.

In the paraxial regime there are three real roots of Eq.~(\ref{eq:2Dray}) in the region inside the caustics and one outside. In the internal region, the lines resulting from demanding constructive/destructive interference of the corresponding rays, Eq.~(\ref{eq:constr}) with $m$ an even/odd integer can be calculated directly. In Fig.~\ref{fig:1} it is shown that, as expected, the lines with $m$ even connect different maximum intensity spots like in \cite{Vaveliuk, maslov,lewis,zapata}. The condition of destructive interference between three waves identified as roots of Eq.~(\ref{eq:AS1Dpxray}) yields optical vortices of order one in the internal region. These points are enclosed by circles illustrated in Fig.~\ref{fig:1}. Notice, however that outside this region there are also optical vortices though there is just one real root of Eq.~(\ref{eq:AS1Dpxray}). This shows the agreement between geometric and undulatory pictures, where constructive/destructive interference of rays inside the caustic region match the light intensity distribution maxima/minima. It also illustrates that the relevant waves whose interference  reproduces optical vortices outside the caustics may not be directly identified by the ray condition.

\subsection{Nonparaxial 2D AS beam}

In a nonparaxial 2D AS beam the coordinates $s_1$ and $s_3$ are scaled with respect to the same parameter $x_{0}$, and the corresponding components of the scaled wavevectors satisfy the exact dispersion relation, $k_{s_{1}}^{2}+k_{s_{3}}^{2}=k_{s}^{2}$. As a consequence, the 2D AS beam spatial function out of the paraxial regime is given by:
\begin{equation}
\Psi_{a_{1}}^{(1)}(s_{1},s_{3})=\frac{1}{2\pi}\int\limits_{-k_{s}}^{+k_{s}}dk_{s_{1}}\mathfrak{S}_{a_{1}}^{(1)}(k_{s_{1}})\mathrm{e}^{i\left(k_{s_{1}}s_{1}+k_{s_{3}}s_{3}\right)},
\label{eq:2DNP} 
\end{equation}
so that the overall phase function is
\begin{equation}
\Phi^{npx(1)}_\mathfrak{AS}(k_{s_{1}};s_1,s_3) =  \vert k_{s_{1}}\vert^{3}/3 - a_{1}^2\vert k_{s_{1}}\vert + k_{s_{1}}s_{1}+k_{s_{3}}s_{3},
\label{eq:2DNPphase}
\end{equation}
where  $ k_{s_{3}}=+\sqrt{k_{s}^{2}-k_{s_{1}}^{2}}$.

\begin{figure}[!h]
\begin{tabular}{cc}
\subfloat[]{\includegraphics[width = 0.24\textwidth]{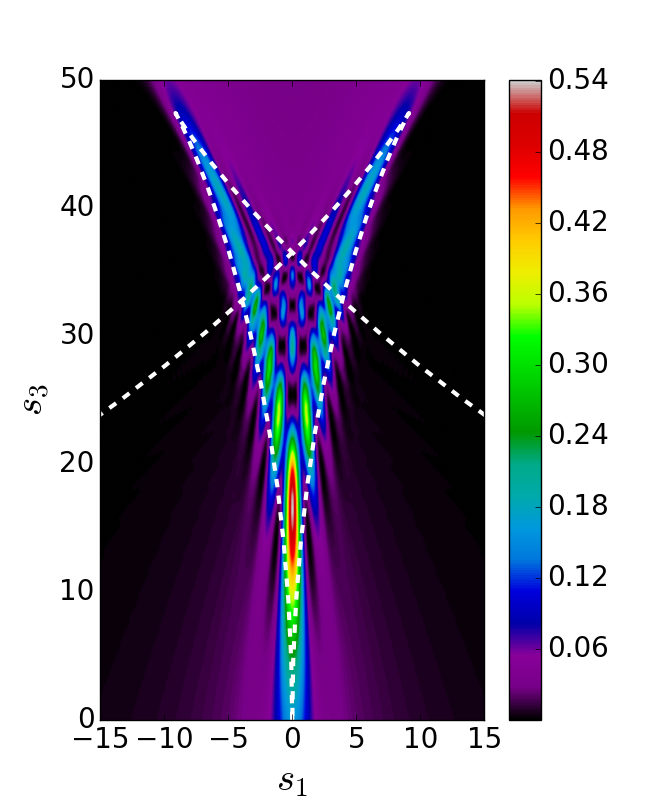}}&
\subfloat[]{\includegraphics[width = 0.24\textwidth]{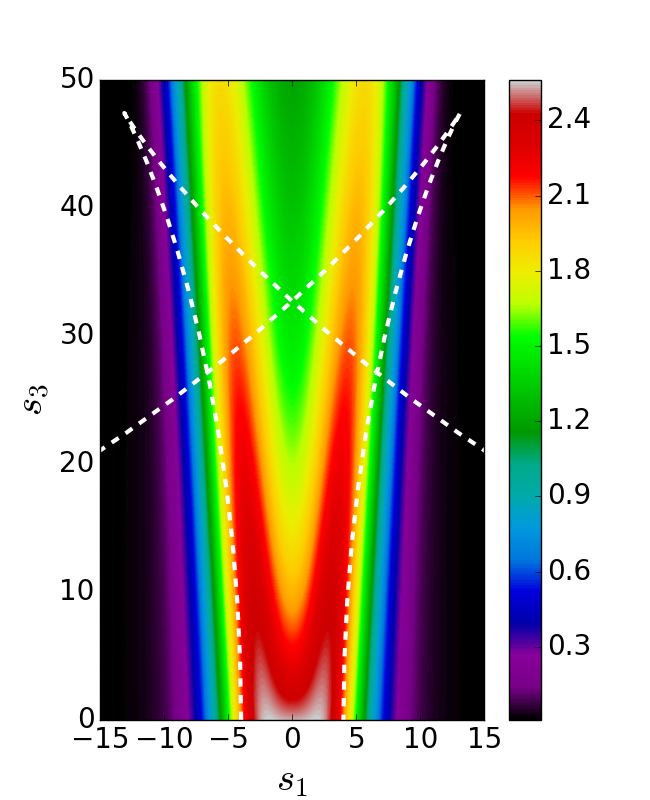}}
\end{tabular}
\caption{Intensity pattern and caustic (white dashed lines) for a 2D nonparaxial AS beam with (a) $a_{1}=0.05$ and (b) $a_{1}=2$. In the first case, $k_{s_1}$ achieves values similar to $k_s=2\pi/0.768 $. In the second case, the value of $a_1$ prevents the formation of a cusp.}
\label{fig:2}
\end{figure}

 Let us study the consequences of a ray description and the corresponding caustic study  out of the paraxial regime for 2D AS beams. Notice that the lines that result from demanding a null value of the first and second derivatives of the phase in  Eq.(\ref{eq:2DNPphase}) with respect to the integration variable $k_{s_1}$ explicitly depend on the frequency of the electromagnetic waves.

Already for paraxial beams, the presence of a cusp nearby $s_1 \sim s_3 \sim 0$ is conditioned to $a_1\ll 1$. Reversing this line of thought, one can say that as the limit $a_1\rightarrow 0$ is approached, a cusp is generated at $s_1 \sim s_3 \sim 0$. This behavior is similar to that arising in the evolution from a Fraunhofer to a Pearcey diffraction pattern \cite{nye4}, in that case increasing smoothly the aperture of a cylindrical lens yields aberrations that manifest as a cusp caustic near the line focus.

Figure \ref{fig:2} illustrates the intensity patterns of two nonparaxial 2D AS beams for two different values of the parameter $a_1$. For $a_1\ll 1$, the intensity pattern resembles that of the paraxial case: it has local maximum intensity spots and a  global maximum intensity spot at the lower cusp, with dark regions nearby. Note, however, that in the nonparaxial regime the number of bright spots is lower. The boundary between the bright spots region and the dark extended region  is delimited by strips of light. This behavior is related to the presence of chains of phase singularities with unitary topological charge. The caustic delimits interesting features of the wave as shown in Fig.~\ref{fig:2}a, although it is not always located close to the most notorious bright regions.

For $a_1 \rightarrow 0$, the caustic exhibits three cusps: one at the origin and two at the top, localized symmetrically about the $s_{3}$ axis. For $a_1\ne 0$ the cusp at the origin is lost: as already mentioned for the paraxial regime, it is replaced by two branches that for $s_3=0$ are separated by  $2a_1^2$, this is illustrated in Fig.~\ref{fig:2}b.

 The ray condition Eq.~(\ref{eq:2Dray}) identifies the set of stationary points $K^{i}_{s_1}$ through the equation
\begin{equation}
(K^{i}_{s_1})^2 + S -\frac{\vert K^{i}_{s_1}\vert s_3}{\sqrt{k_s^2 -(K^{i}_{s_1})^2}}=0,\quad\quad S = \mathrm{sign}(K^{i}_{s_1})s_1 - a_1^2
\label{eq:orroots} 
\end{equation}
for nonparaxial  2D AS beams. This expression can be accomplished only if  the 6th-order polynomial equation,
\begin{eqnarray}
a(K^{i}_{s_1})^6 + b(K^{i}_{s_1})^4 + c (K^{i}_{s_1})^2 &+& d= 0,
\label{eq:effroots}\\
a= -1,\quad b = k_s^2 - 2S,&&\nonumber\\ c = 2k_s^2S- S^2 - s_3^2,\quad d = k_s^2S^2 &,&\nonumber
\end{eqnarray}
is fulfilled. Taking as variable the term $(K^{i}_{s_1})^2$, there exist three different real roots of this equation whenever the discriminant
\begin{equation}
\Delta =18abcd-4b^{3}d+b^{2}c^{2}-4ac^{3}-27a^{2}d^{2}
\end{equation}
is greater than zero. For $\Delta = 0$ the roots are real but a degeneracy arises, and for $\Delta<0$ the equation has one real root and two non-real complex conjugate roots. In the particular case $S=0$, there are five roots 
\begin{eqnarray}
&& K^{1}_{s_1} =0,\quad K^{(2,3)}_{s_1} =\pm\sqrt{\frac{k_s^2}{2} - \sqrt{
\Big(\frac{k_s^2}{2}\Big)^2 - s_3^2}}, \\ \nonumber
&&
\quad
 K^{(4,5)}_{s_1} =\pm\sqrt{\frac{k_s^2}{2} + \sqrt{
\Big(\frac{k_s^2}{2}\Big)^2 - s_3^2}}.
\end{eqnarray}
The root $K^{1}_{s_1} =0$ should be understood as a limit $K^{1}_{s_1} \rightarrow 0$ from both positive and negative neighbourhoods yielding rays with the equation $\vert s_1\vert = a_1^2$.

Once the roots of Eq.~(\ref{eq:effroots}) are identified, it must be verified that they satisfy both the dispersion relation and the original Eq.~(\ref{eq:orroots}).

\begin{figure}[!h]
\begin{tabular}{cc}
\subfloat{\includegraphics[width = 1.8in]{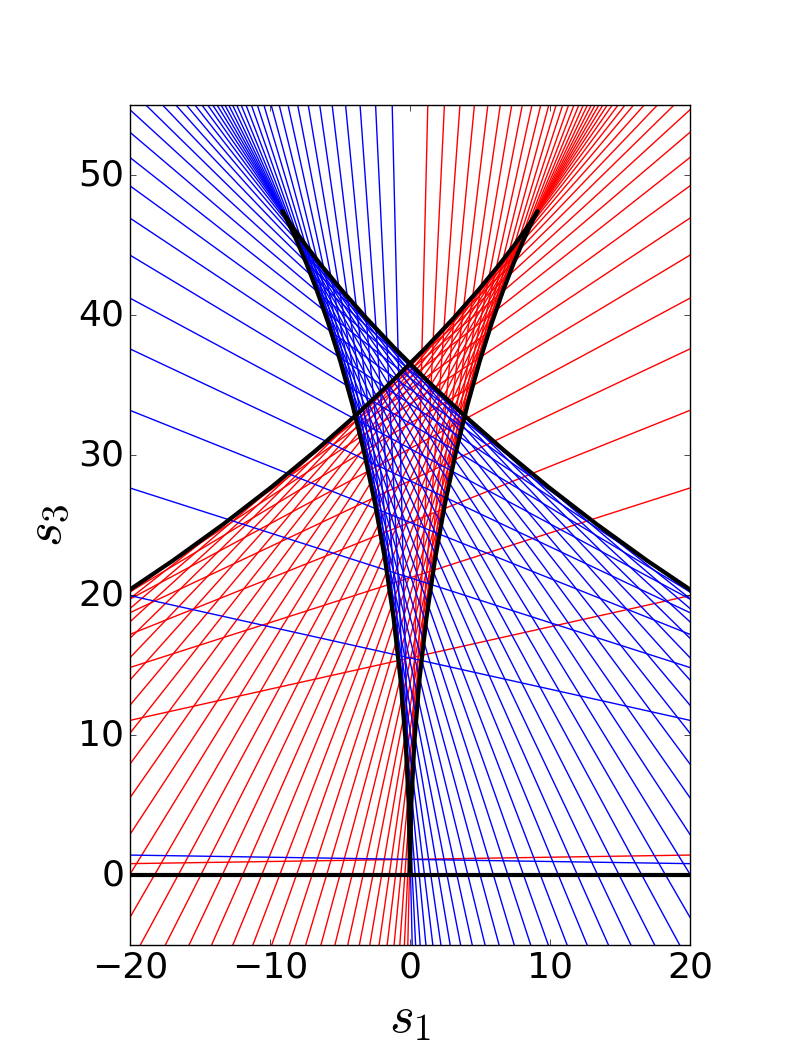}}&
\subfloat{\includegraphics[width = 1.8in] {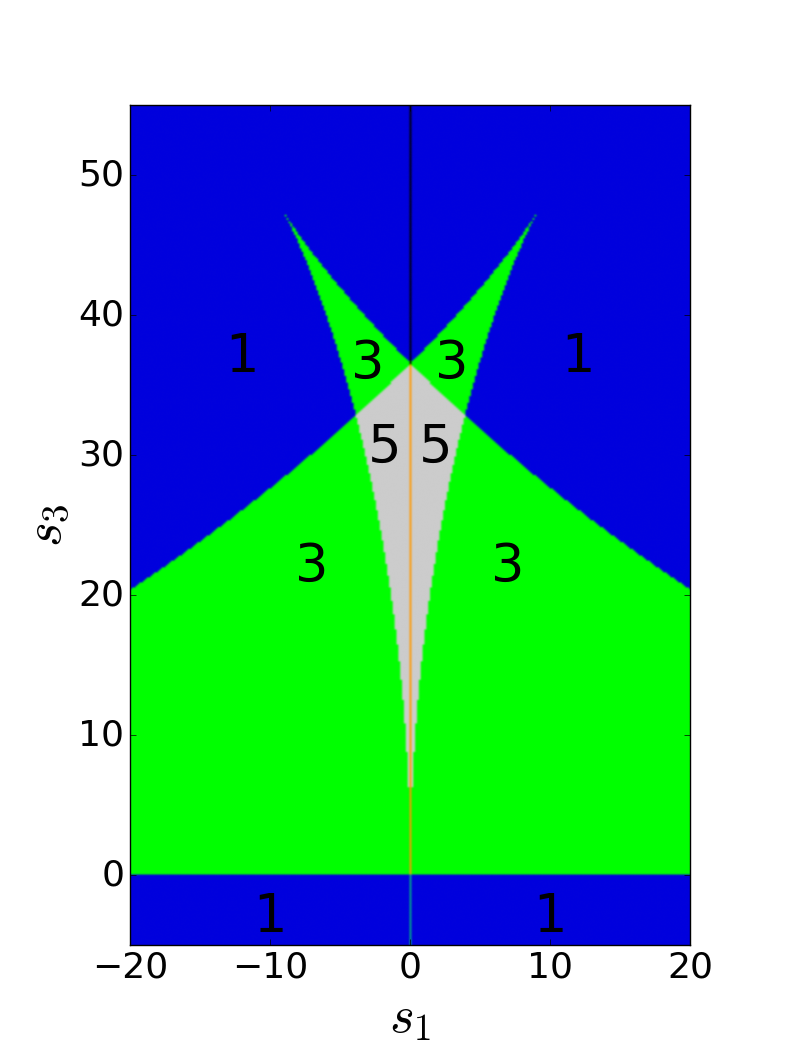}}\\
(a) & (b)\\
\subfloat{\includegraphics[width = 1.8in]{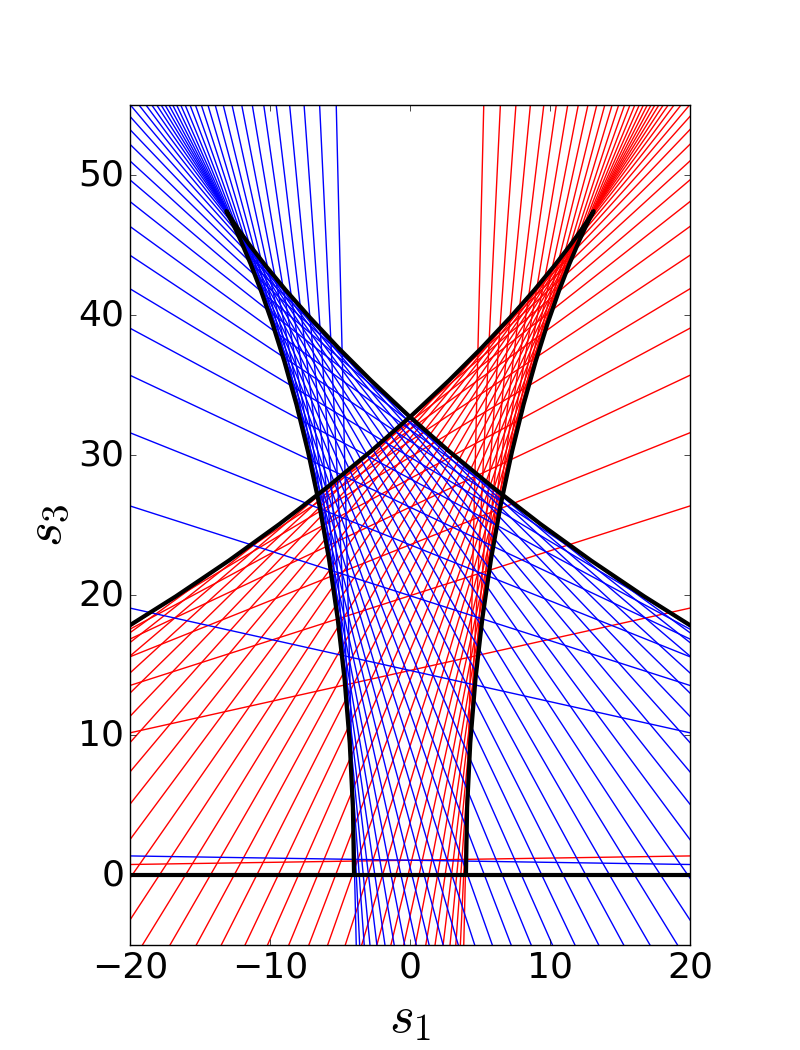}}&
\subfloat{\includegraphics[width = 1.8in]{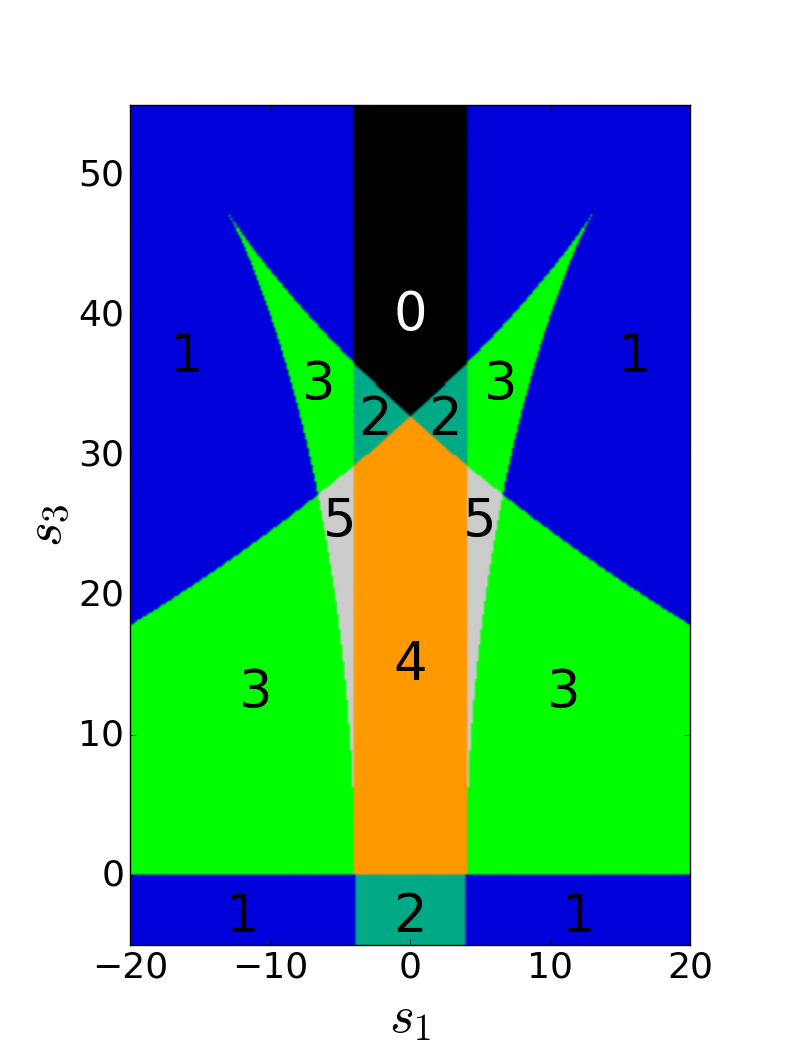}}\\
(c) & (d)\\
\end{tabular}
\caption{(a,c) Caustic (black lines) and rays, (b,d) number of relevant roots of the ray condition for a 2D nonparaxial AS beam with $a_{1}=0.05,2$ and $k_s=2\pi/0.768 $. Red and blue lines illustrate the rays obtained by the positive and negative components of the roots $K_{s_{1}}$, respectively, taken within the interval ($-k_{s},+k_{s})$. The root $K_{s_{1}}=0$ is not considered.}
\label{fig:2DNPcau}
\end{figure}

Fig.~\ref{fig:2DNPcau} illustrates caustics, rays and number of relevant roots of the ray condition of 2D AS nonparaxial beams.  While the  root number for paraxial 2D AS beams is constrained to be either three or one, for nonparaxial AS beams, the number of different roots that give $K_{s_1}$ values that can be interpreted as components of the wave vector $\mathbf{k}_s$ is also space dependent but with values  in the set $\{0,1,2,3,4,5\}$. That is, there is no finite region where six relevant nondegenerate roots of Eq.(\ref{eq:effroots}) could be identified. The expected discontinuity in the number of available real roots is observed when crossing the caustic. Notice also that the difference in the number  of roots $K_{s_1}^{i}$ when crossing the boundary of the region $s_1\in(-a_1^2, a_1^2)$ towards the region  $\vert s_1\vert >a_1^2$ is just one.

\begin{figure}[!h]
\begin{tabular}{cc}
\subfloat{\includegraphics[width = 1.8in, trim = {4cm 0cm 4cm 0cm}, clip=true]{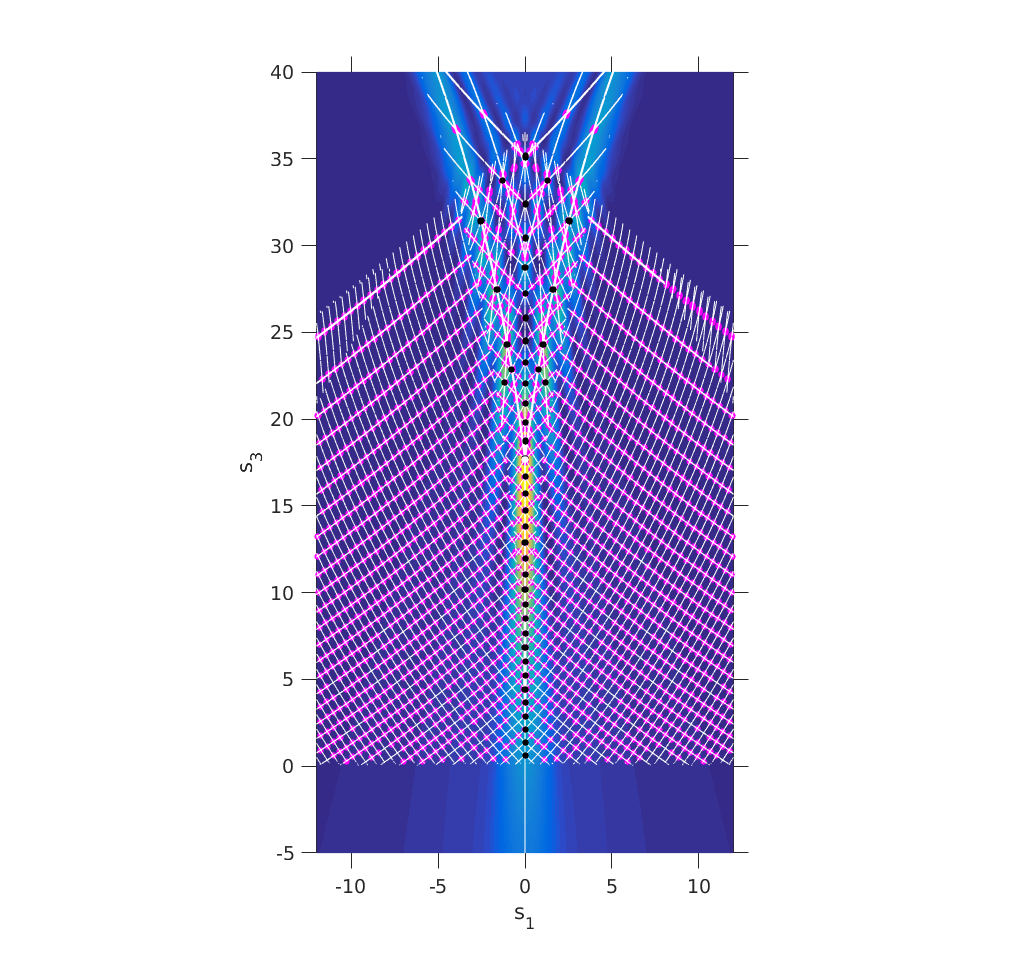}}&
\subfloat{\includegraphics[width = 1.8in, trim = {4cm 0cm 4cm 0cm}, clip=true] {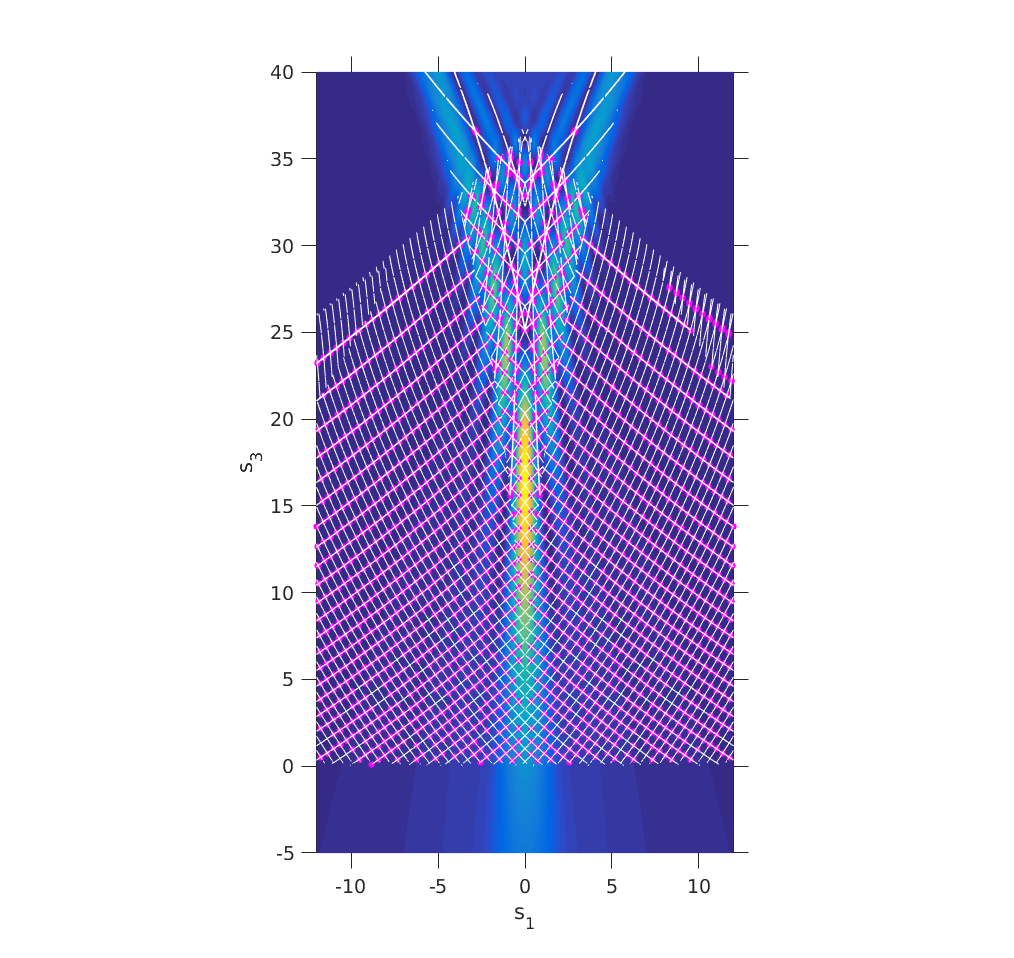}}\\
(a) & (b)\\
\subfloat{\includegraphics[width = 1.8in, trim = {4cm 0cm 4cm 0cm}, clip=true]{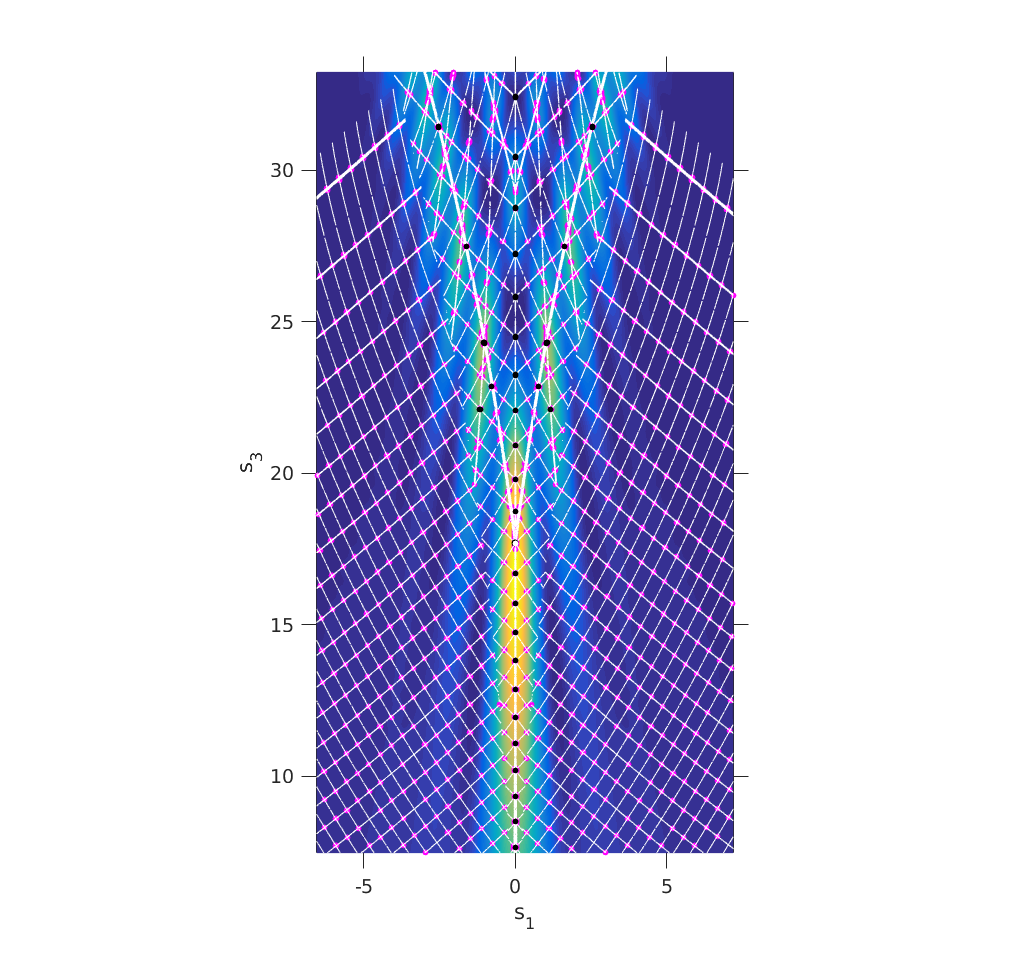}}&
\subfloat{\includegraphics[width = 1.8in, trim = {4cm 0cm 4cm 0cm}, clip=true]{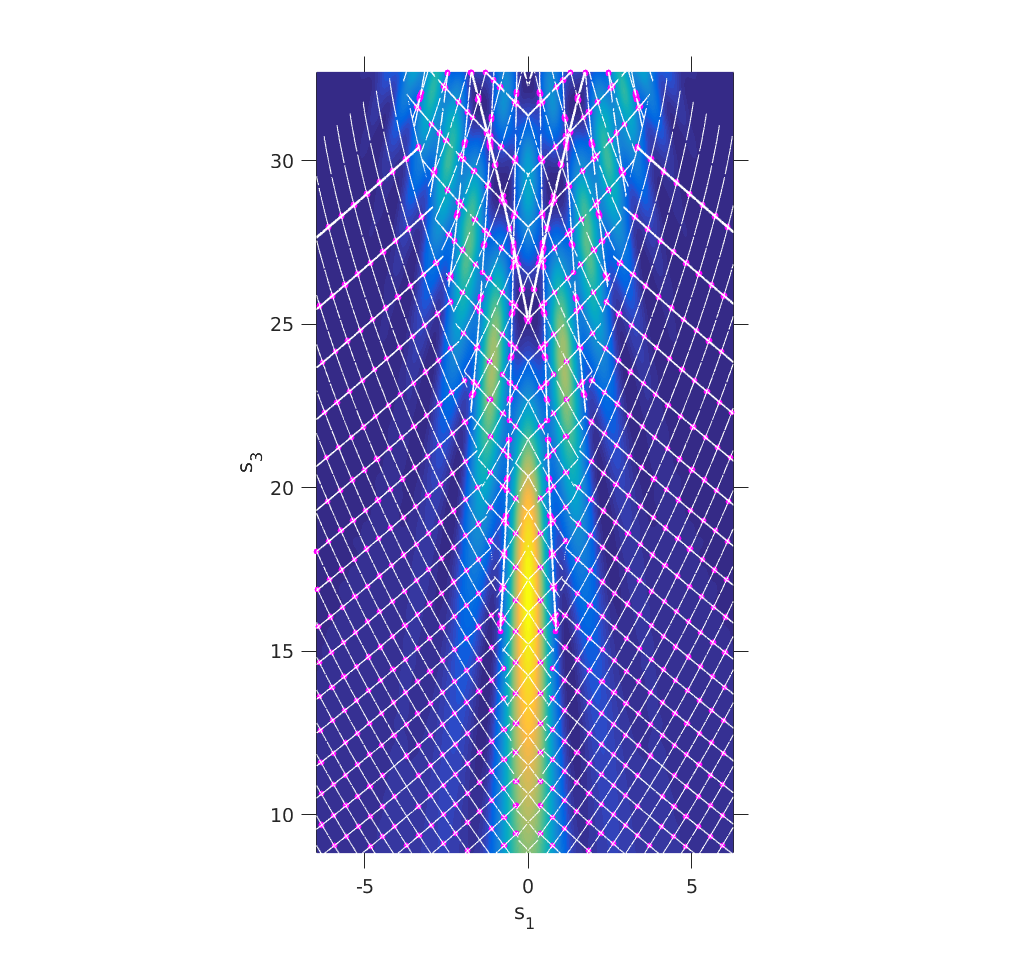}}\\
(c) & (d)\\
\end{tabular}
\caption{Results of applying the algorithm described in Section II.A to find lines that join (a) high intensity points and (b) vortices for a nonparaxial 2D AS beam given by Eq.~(\ref{eq:2DNP}) with $a_1 =0.05$ and $k_s= 2\pi/0.768$, (c) and (d) are enlarged sections of (a) and (b), respectively. The accumulated number of constructive interferences is marked with different colors, white (I=+1), magenta (I=+2), black (I=+3); while destructive interference corresponds to white (I=-1) and magenta (I=-2).}
\label{fig:linE1}
\end{figure}

Similarly to the paraxial case, the  stationary condition that defines the rays, Eq.~(\ref{eq:2Dray}), can be used to understand the mathematical features behind the location of the spots where the wave amplitude takes local maximum and minimum values.
The numerical algorithm described in section III is then applied for nonparaxial AS beams. In this case, there are more roots $K^i_{s_1}$ than in the paraxial case. Numerical simulations reveal that for 2D AS nonparaxial beams, close to a maximum, there are always two roots with similar amplitude that is an order of magnitude higher than the amplitude of the  other roots. For $(s_1,s_3)$ close to a vortex there is just one root with an amplitude twice or higher value than the amplitude of the other roots. Neglecting the relative amplitudes with a smaller value than $10^{-3}$ we obtain the results shown in Fig.~\ref{fig:linE1}. There it is shown that the resulting lines constructed for the conditions of constructive (destructive) interference of the waves with non negligible amplitude locate the maximum (minimum) amplitudes in the field. Nevertheless, optical vortices inside the caustic involve three waves with just two of comparable absolute value of their amplitude.

\begin{figure}[!h]
\begin{tabular}{cc}
\subfloat{\includegraphics[width = 1.8in, trim = {4cm 0cm 4cm 0cm}, clip=true]{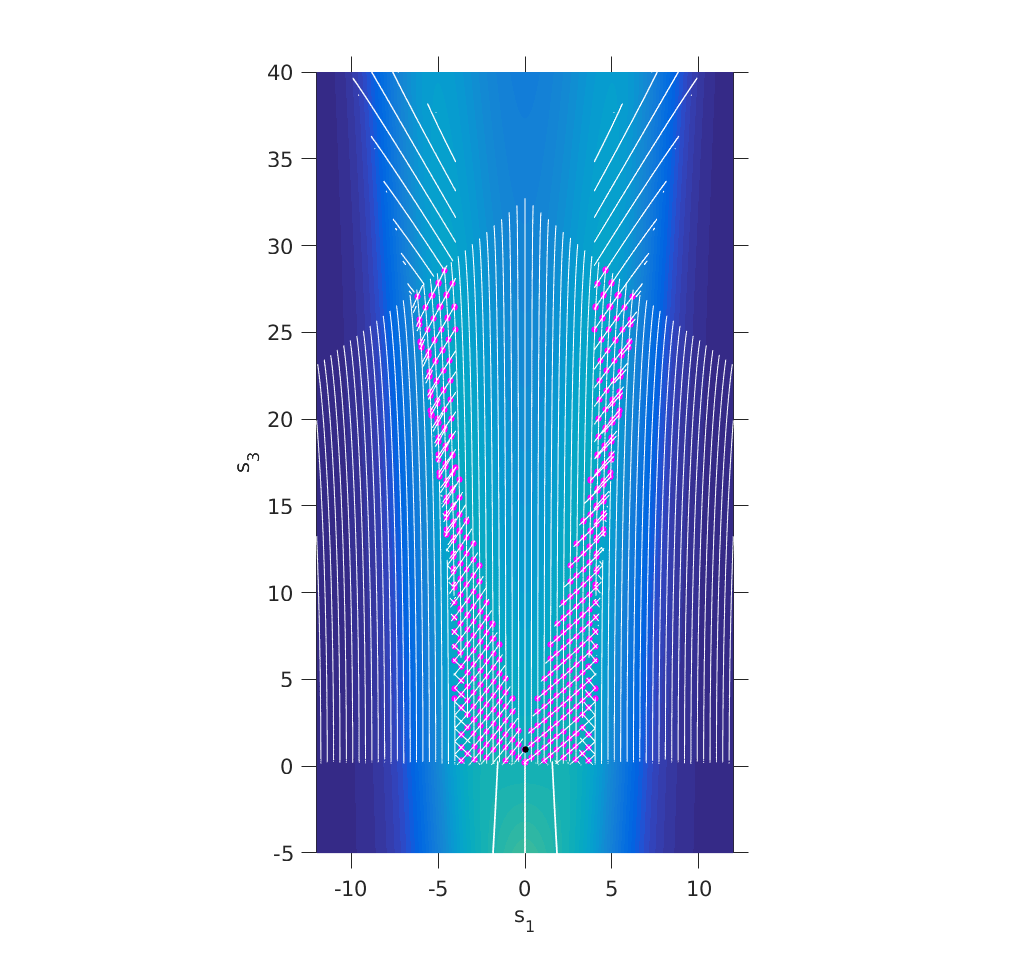}}&
\subfloat{\includegraphics[width = 1.8in, trim = {4cm 0cm 4cm 0cm}, clip=true] {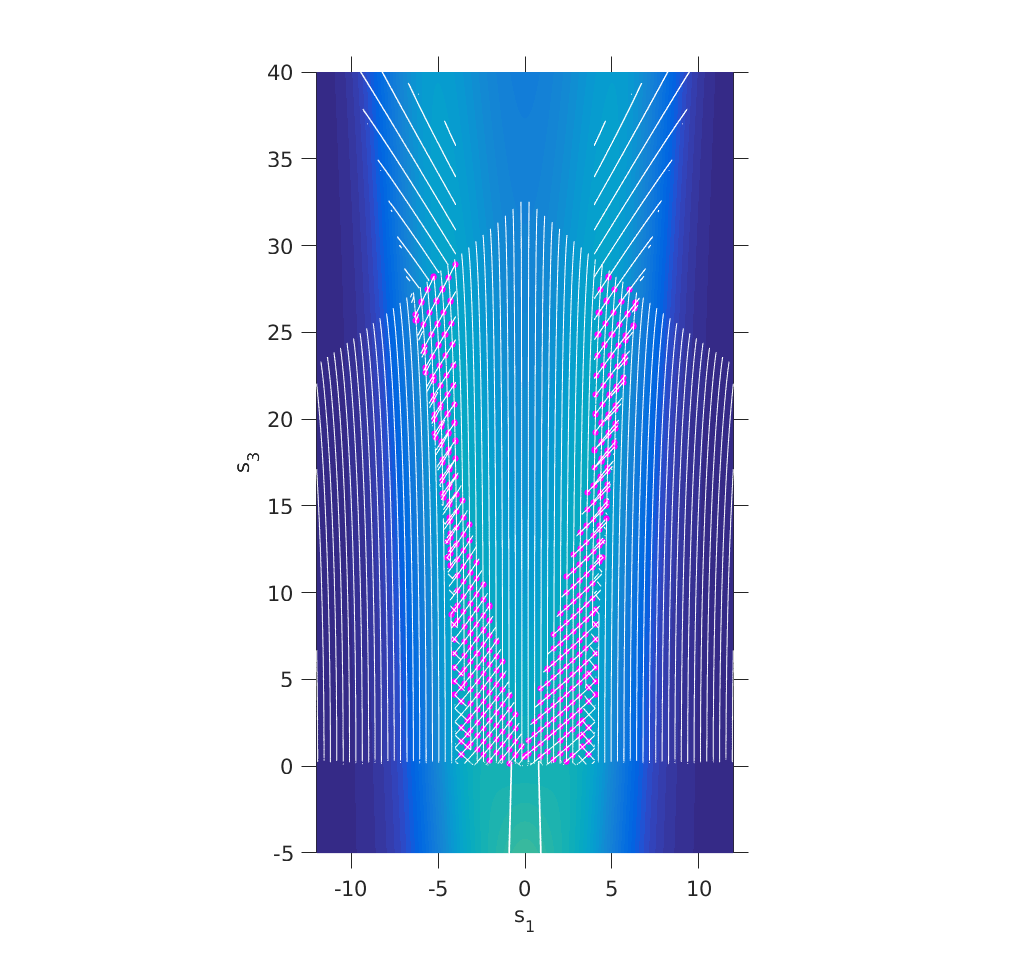}}\\
(a) & (b)\\
\end{tabular}
\caption{Results of applying the algorithm described in Section II.A to find lines that join (a) high intensity points and (b) vortices for a nonparaxial 2D AS beam given by Eq.~(\ref{eq:2DNP}) with $a_1 =2$ and $k_s= 2\pi/0.768$. The accumulated number of constructive interferences is marked with different colors, white (I=+1), magenta (I=+2), black (I=+3); while destructive interference corresponds to white (I=-1) and magenta (I=-2).}
\label{fig:2DNPa2lin}
\end{figure}

For greater values of the parameter $a_1$, the number of rays define six different regions as illustrated in Fig.~\ref{fig:2DNPcau}d. The results of applying the algorithm to identify maxima and minima
is shown in Fig.~\ref{fig:2DNPa2lin}

\subsection{Morphology of the  component of the electric field along the main propagation axis $E_{s_{3}}$}

Following the construction of electromagnetic beams described in Section II.A, Eqs.~(22 -23), the electric field has a component along the $s_3$ direction that can be obtained from the electric field  component along the $s_1$ direction, $E_{s_{1}}$. The fulfillment of Gauss equation in a medium without free charges requires that the former is given by:
\begin{equation}
E_{s_{3}}(s_{1},s_{3})=-\frac{1}{2\pi}\int\limits_{-k_{s}}^{k_{s}}dk_{s_{1}}\frac{k_{s_{1}}}{k_{s_{3}}}\mathfrak{S}_{a_{1}}^{(1)}(k_{s_{1}})\mathrm{e}^{i\left(\mathbf{k}_{s}\cdot\mathbf{s}\right)}.
\end{equation}

Fig.~6 illustrates the intensity patterns of the $E_{s_{3}}$ component for $a_1\ll 1$. The amplitude of $E_{s_{3}}$ is antisymmetric with respect to reflections on the $s_3$ axis, so that the intensity pattern of $E_{s_{3}}$ is symmetric with a null value at $s_1 = 0$. The maximum intensities that it can reach  are smaller than  those for the $E_{s_{1}}$ component (about five percent in this illustrative example). $\vert E_{s_{3}}\vert^2$ exhibits two equal maximum spots at the {\bf top} of the perceptible region of the beam and no bright spot at the {\bf bottom}. In fact $E_{s_{3}}$ interchanges the regions with strips of local maximum intensity in the $E_{s_1}$ field with dark regions. In the phase pattern there is a discontinuity over the $s_{3}$ axis, \textit{i.e.} for $s_{1}=0$, and this is also due to the parity of the integrand in $E_{s_{3}}$. It is still possible to find at the sides chains of phase singularities similar to those in $E_{s_{1}}$, as well as singularities around the center distributed symmetrically with respect to $s_{3}$ axis. The latter have opposite unitary topological charge.

The rays and caustic for $E_{s_{3}}$ are the same than those of $E_{s_{1}}$  because they share, up to a constant, the same phase function in ${\bf k}$ space: the amplitude factor $k_{s_{1}}/k_{s_{3}}$ add an additional phase factor of $\pi$ for negative critical $K_{s_1}/K_{s_3}$ satisfying the rays condition. Since constructive or destructive interference results from the $\{\Delta\Phi^{\ell,\ell^\prime}_{i},i=1,...N\}$  relations described in last subsection, the component $E_{s_{3}}$ of the 2D AS EM wave exhibits maxima and minima intensity spots that are mostly interchanged with respect to the $E_{s_1}$ component, as Fig~\ref{fig:linE3} illustrates. Notice that the amplitude factor $k_{s_{1}}/k_{s_{3}}$ also modifies the relative amplitude $A_\Delta (K^M_{s_1}, K^i_{s_1})$ with respect to that of the $E_{s_1}$ field. This has consequences on the relevance of any given ray in the approximate description of the beam given by Eq.(\ref{eq:rel}).

\begin{figure}[!h]
\includegraphics[width=6cm]{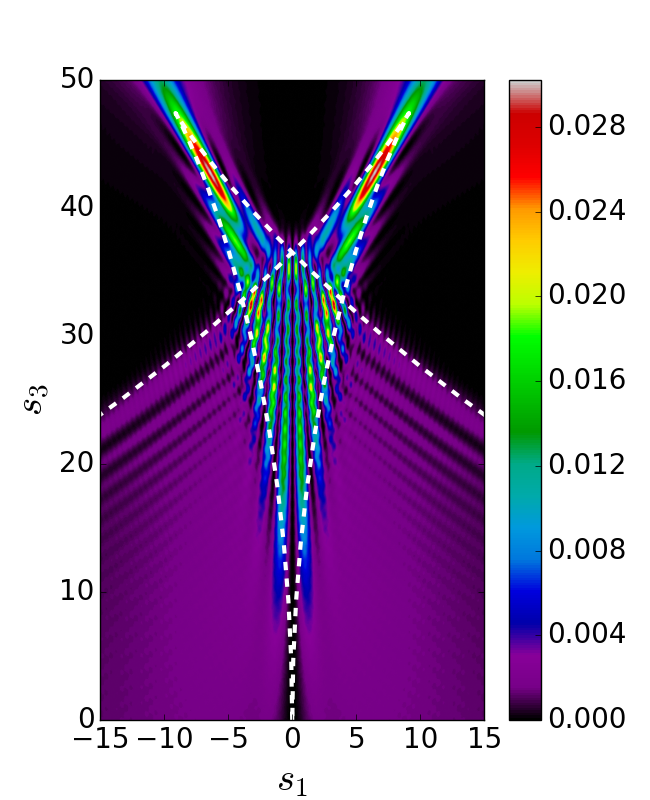}
\caption{Intensity pattern and caustic (white dashed line) for the $E_{s_{3}}$ component of the 2D nonparaxial AS vectorial beam obtained with $a_{1}=0.05$ and $k_s=2\pi/0.768$.}
\label{fig:Lcomp}
\end{figure}

\begin{figure}[!h]
\begin{tabular}{cc}
\subfloat{\includegraphics[width = 1.8in, trim = {4cm 0cm 4cm 0cm}, clip=true]{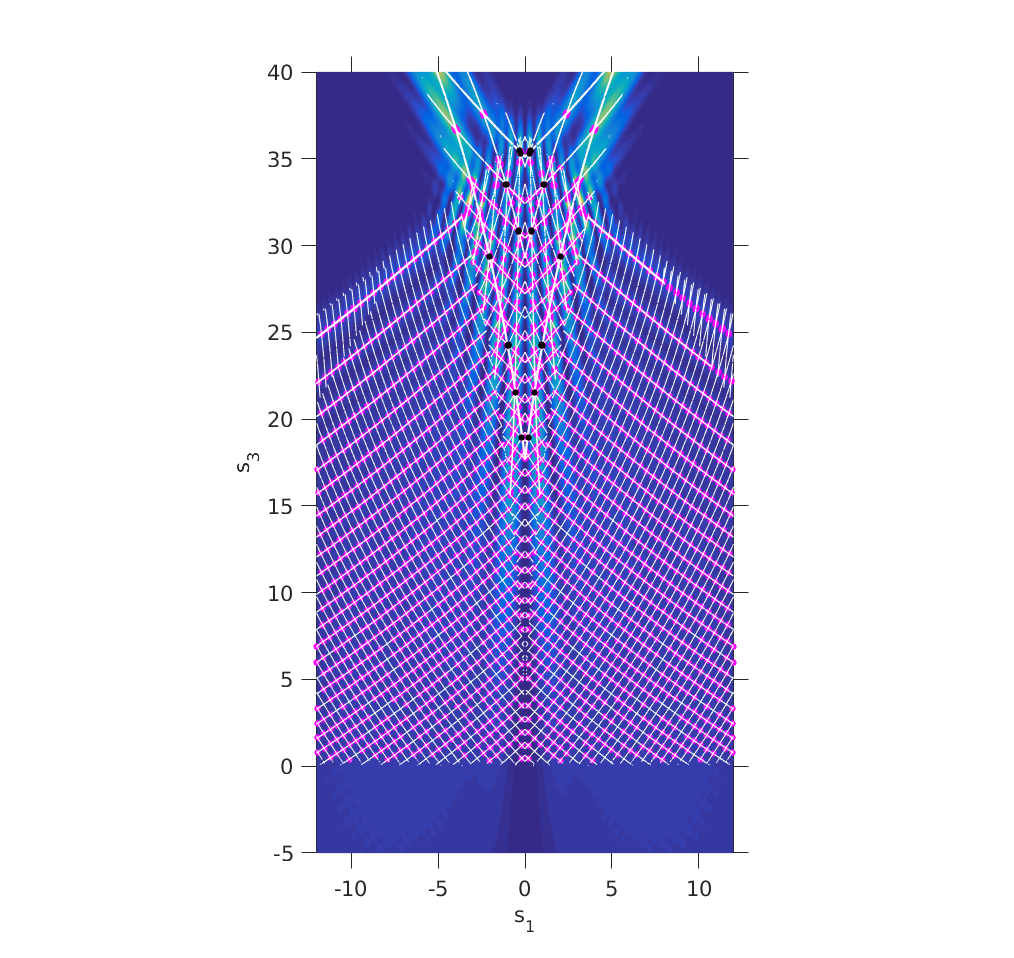}}&
\subfloat{\includegraphics[width = 1.8in, trim = {4cm 0cm 4cm 0cm}, clip=true]{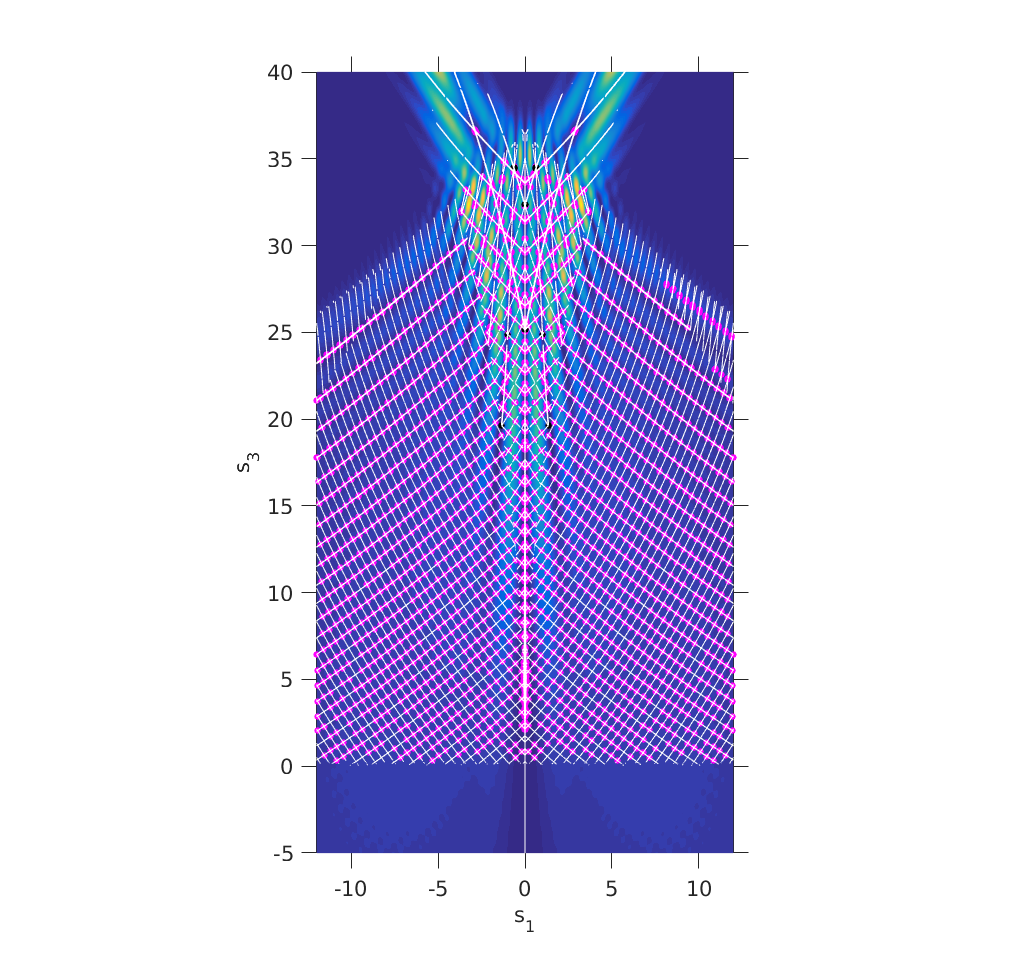}}\\
(a) & (b)\\
\subfloat{\includegraphics[width = 1.8in, trim = {4cm 0cm 4cm 0cm}, clip=true]{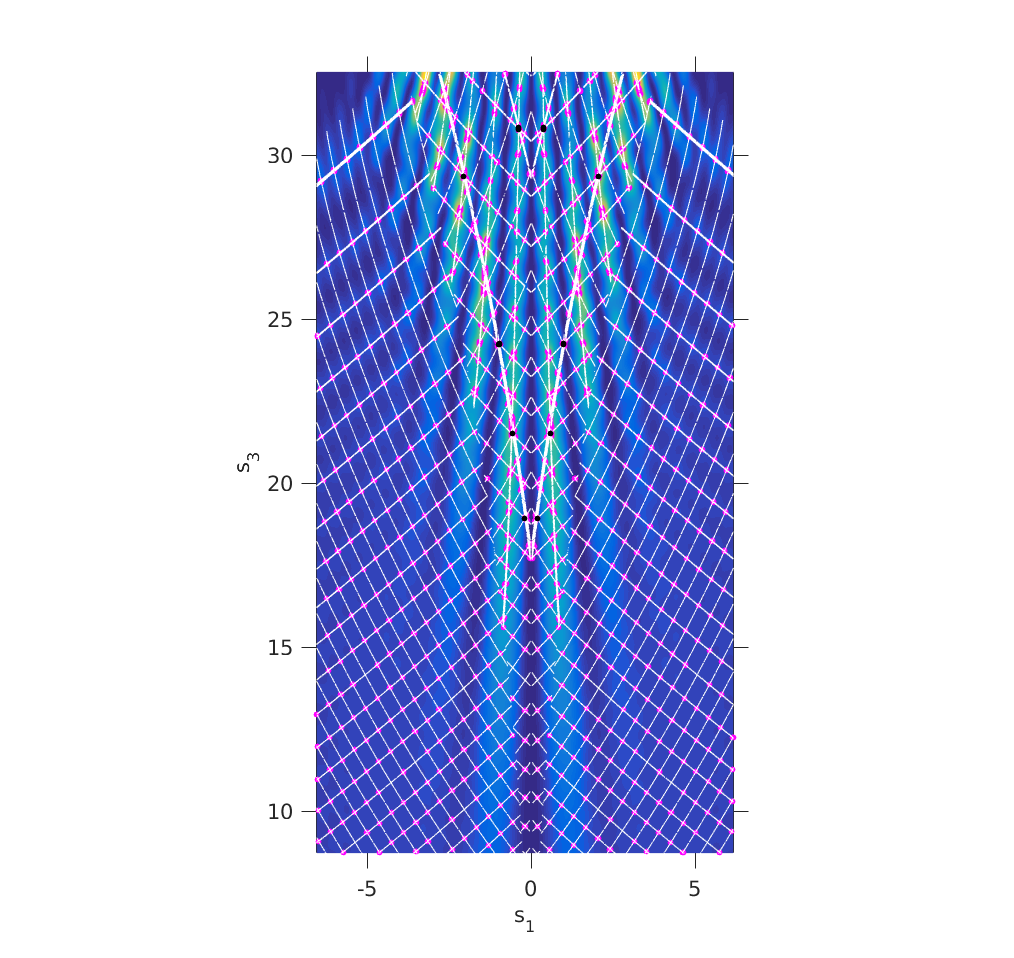}}&
\subfloat{\includegraphics[width = 1.8in, trim = {4cm 0cm 4cm 0cm}, clip=true]{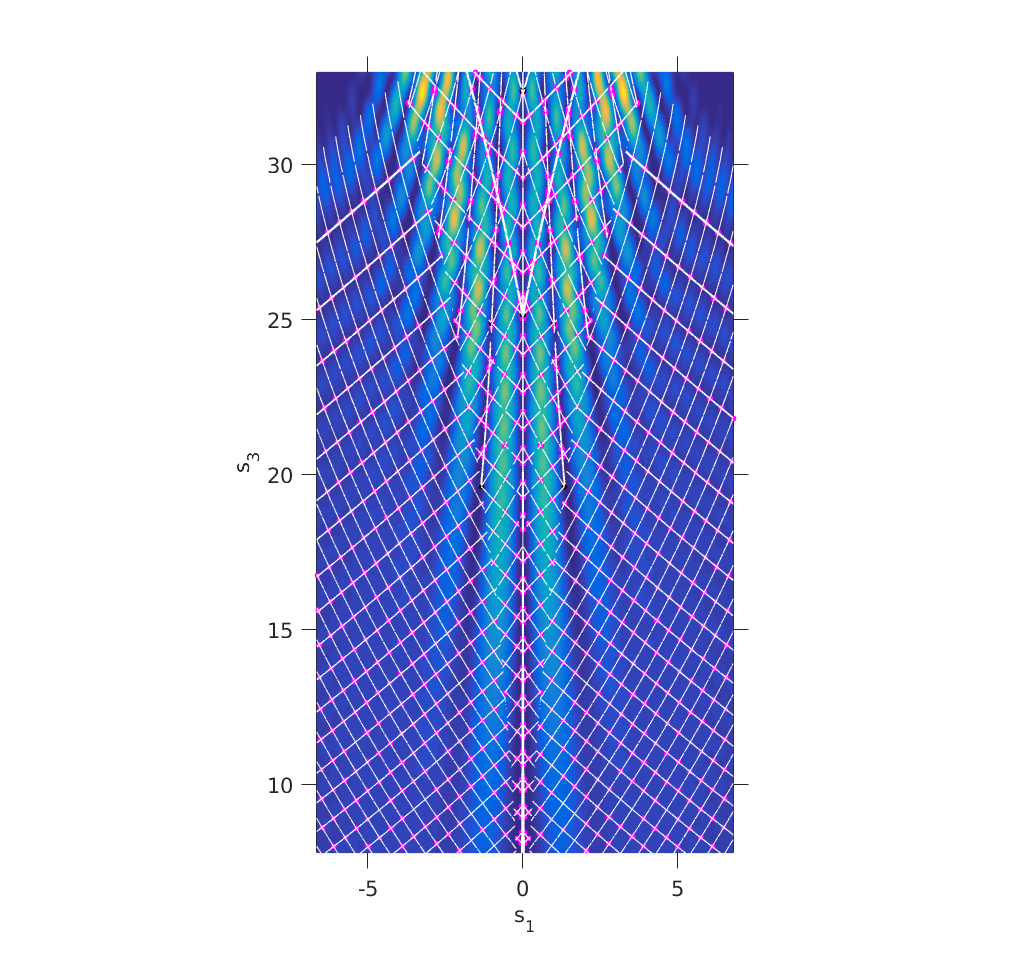}}\\
(c) & (d)\\
\end{tabular}
\caption{Results of applying the algorithm described in Section II.A to find lines that join (a) high intensity points and (b) vortices for the longitudinal component of a nonparaxial 2D AS beam with $a_1 =0.05$ and $k_s= 2\pi/0.768$, (c) and (d) are enlarged sections of (a) and (b), respectively. The accumulated number of constructive interferences is marked with different colors, white (I=+1), magenta (I=+2), black (I=+3); while destructive interference corresponds to white (I=-1) and magenta (I=-2).}
\label{fig:linE3}
\end{figure}

\section{Conclusions.}

A careful analysis of the morphology of Airy symmetric beams has been performed. The morphology of 2D AS beams is similar to that expected for a cusp catastrophe, whenever the parameter $a_1$ introduced in their definition is small, $a_1\ll 1$; the resemblance is more evident for paraxial 2D AS beams. This occurs in spite of the third  power of the potential function of AS beams compared to the fourth power of Pearcey beams; the similarity is a consequence of the even parity under reflection of $k_{s_1}$ that the angular spectra of 2D AS beams and Pearcey beams share. As $a_1$ increases, 2D AS  beams lose the cusp and the number of phase singularities diminishes. In this regime, the intensity pattern tends to be dominated by a single high intensity region.

We have shown the relevance of supplementing  an undulatory analysis using geometric ideas to have a better understanding of electromagnetic fields. Our proposed algorithm extends previous studies that allow the determination of curves that join different local maxima and minima of the intensity pattern of structured light. Those algorithms were previously tested for 2D beams in the paraxial regime. Our algorithm recovers the results obtained before, and has shown to be useful in the non paraxial regime. There, in general, critical points satisfy equations that differ structurally from their paraxial analogues. We applied that algorithm for 2D AS beams recognizing that the parameter $a_1$ introduced originally for a description of waves with a finite energy introduces new and interesting features. The parameter $a_1$ both determines the relevant values of  $k_{s_1}$ through an effective Gaussian envelope, and play a central role in the derivatives of the overall phase.
Non negligible values of $a_1$ yield caustics separated in two branches, and the spatial distribution of the number of rays acquires a richer structure. 
 
So far, an experimental realization of nonparaxial AS beams with $a_1 >1$ has not been reported. However, our results suggest that 2D AS beams can be used to study experimentally the formation of optical singularities as a function of the $a_1$ parameter. We  have also found that for AS 2D electromagnetic waves, the morphology of the longitudinal electric field component exhibits complementary features with respect to the transverse component. This complementarity manifests in an interchange between local maximum regions with regions containing a phase singularity. It would be important to analyze whether or not this phenomenon is observable for other electromagnetic fields with  longitudinal electric component satisfying  Eq.~(\ref{eq:vector}).

We have also found some limitations of the ray equation for the identification of plane waves whose superposition leads to the formation of some optical singularities. That is exemplified by
the presence of just one root of the ray equation in the region outside the caustics in Fig.~\ref{fig:1}, and the existence of optical vortices located there. We should also remark that our algorithm locates local maxima and minima according to the stationary phase approximation to the exact expression of a structured beam. The relative intensity of those local maxima with respect to a global maximum is not discerned from our algorithm. This property is behind the presence of critical points with a parameter $I=+3$ in Fig.~(\ref{fig:linE3}) which, however do not correspond to regions of the highest intensity.

Our results also suggest that the different singularities in paraxial and nonparaxial AS beams could have interesting effects when interacting with material particles like atoms, nano and micro structures. The response could be highly sensitive on the parameters of the electromagnetic  AS beam.

Finally, future work could include studying the effects of evanescent waves, which are expected to be particularly interesting for non-paraxial electromagnetic waves propagating, e. g.,  in media with space dependent refractive indexes. In this context, the relevance of the ray equation for a selection of  co-moving frames could be explored. Those frames have already produced very interesting results in moving \cite{leonhart99} or non-homogeneous media \cite{aiello2013}.

\begin{acknowledgements}
This work was partially supported by CONACyT LN-293471, DGAPA  IN107719. F.C.A. acknowledges a CONACyT scholarship.
\end{acknowledgements}

\end{document}